# α-铁中高能级联损伤研究

王炜路[1]，刘伟[1]，刘长松[1*]，方前锋[1]，黄群英[2]，吴宜灿[2]

[1]中国科学院固体物理研究所材料物理重点实验室，安徽 合肥，230031

[2]中国科学院等离子体物理研究所，安徽 合肥，230031

级联能高达 250 keV 的 α-铁级联碰撞分子动力学模拟被研究，相当于实现了中子能量为 14.1 MeV 的辐照损伤模拟。考虑到统计的不确定性，弗兰克尔缺陷对产生效率和完美晶格损伤效率随着级联能的增加有相近的趋势。这两个量先按照幂律的形式呈下降趋势，当子级联能达到子级联阈能值时达到最小，然后轻微地呈上升趋势。近一步的分析展示，以上两个量随级联能的行为与缺陷的分散程度有紧密的关联，这两个量都取决于子级联的数量和能量。在本文中对分析技术首次被用于级联碰撞的研究，得到了令人满意的研究结果。

**关键词**：α-铁，分子动力学模拟，级联碰撞，对分析技术

作者简介：

王炜路（1980- ），男，中国科学院固体物理研究所博士生。目前主要从事中子辐照效应分子动力学模拟研究。E-mail: wlwang@issp.ac.cn。

# Cascade damage in α-iron with high damage energy

WANG Weilu (王炜路)[1], LIU Wei (刘伟)[1], LIU Changsong (刘长松)[1,*], FANG Qianfeng (方前峰)[1], HUANG Qunying (黄群英)[2], WU Yican (吴宜灿)[2]

[1] Key Laboratory of Materials Physics, Institute of Solid State Physics, Chinese Academy of Sciences, Hefei 230031, China

[2] Institute of Plasma Physics, Chinese Academy of Sciences, Hefei 230031, China

**Abstract**

A molecular dynamics study was done on cascades damage in α-iron with high cascade energy up to 250 keV, which is corresponding to neutron energy above 14.1 MeV. We observed that Frenkel pairs production efficiency and perfect crystal lattices damage efficiency have similar trend within statistical uncertainty with the increment of cascade energy: first decrease according to power-law and arrive at minima at a threshold energy of subcascades, then increase in a narrow range. Further analysis demonstrates that behavior of the above two quantities versus cascade energy has close relationship with defect dispersion which depend on the number and energy of subcascades. In this paper pair analysis technique is applied in cascades damage for the first time and its efficiency is well justified.

* supported by the Innovation Program of Chinese Academy of Sciences (Grant No.: KJCX2-YW-N35) and the National Magnetic Confinement Fusion Program (Grant No.: 2009GB106005)

## 1. Introduction

Neutron irradiation damage has received increasing interest because of the need to assess structural material performance in first generation deuterium-tritium (DT) fusion reactors such as ITER. However, it is difficult to conduct irradiation experiments in which structural materials undergo high-energy neutrons up to 14.1 MeV. The increase in computer power because of modern machines and the development of many-body potentials has resulted in an opportunity to use molecular dynamics (MD) to perform an in-depth examination of cascade evolution with a lifetime of only a few picoseconds and a damage zone with a maximum volume of $10^7$ to $10^8$ atoms. Despite progress in this field, the physics of cascade damage remains interesting, and new behaviors or effects could still be discovered and investigated.

Since the first paper on MD simulation of displacement cascades in α-iron [1] using the first Finnis-Sinclair many-body potential [2,3], stiffened by Calder and Bacon to treat small interatomic distances properly [1], many MD simulations of displacement cascades in α-iron have been performed [4-7]. Malerba (2006) [4] indicated that discrepancies in the criterion used to define point-defects and clusters are largely responsible for the different results on the number of Frenkel pairs ($N_F$) and point-defect clustered fractions. Recently, Björkas (2007) [5] used three different interatomic potentials with finer descriptions of interstitial energy to simulate the displacement cascade with MD, and discovered that the total Frenkel pair production is the same within the statistical uncertainty. However differences remain in the fraction of clustered defects. While analyzing about Frenkel pairs production efficiency (FPPE) or the number of Frenkel pairs produced per cascade ($\eta_{FP}$) that survived in-cascade recombination at the end of the simulations, Bacon (1995) [6] proposed a formula

$$\eta_{FP} = N_F / \nu_{NRT} = N_F / (0.8 E_{MD} / 2 E_d) = 0.567 E_{MD}^{(-0.221)} \qquad (1)$$

with cascade energy up to 20 keV at 100 K. Here, $\nu_{NRT}$ is the number of Norton-Robinson-Torrens (NRT) displacements that is the number of Frenkel pairs generated by a primary knock-on of initial kinetic energy or damage energy ($E_{MD}$) with the classical linear theory NRT law [7,8]. Generally, when electronic excitation is not included, it is considered coincident with the cascade energy in the MD simulation, which has a relationship with the energy of primary knockon atoms (PKA) ($E_{PKA}$) and neutron energy [9]. When neutron energy is 14.1 MeV, the corresponding $E_{PKA}$ is 487 keV and $E_{MD}$ is 220.4 keV. $E_d$ is the average threshold displacement energy (TDE) [10] and is considered equal to 40 eV. Stoller (1999) [9] gave a function

$$\eta_{FP} = 0.5608 E_{MD}^{(-0.3029)} + 3.227 \times 10^{-3} E_{MD} \qquad (2)$$

with cascade energy up to 40 keV at 100, 600 and 900 K. In the equation, the first term dominates the energy dependence up to about 20 keV. The second term is responsible for the minimum in the FPPE curve at about 20 keV. This change in energy dependence occurs because subcascade formation makes a single high-energy cascade appear to be the equivalent of several lower energy cascades. Thus, the FPPE is slightly higher at 40 than at 20 keV. It appears unlikely that $\eta_{FP}$ will change significantly at higher cascade energies. Experimental [11-16] studies have confirmed that the FPPE agrees with experimental estimates up to 100 keV.

In this paper, using the potential by Ackland, Mendelev and Srolovitz et al. (AMS) [17], which describes the interstitial energy well, a statistically significant number of simulations were carried out at cascade energies of 0.5-250 keV and temperatures of 100 and 300 K for the comparison. For making up for the information of defects due to discrepancies of criterion, the pair analysis technique [18-19] was performed to study cascades damage, which may help us understand perfect crystal lattices damage efficiency (PCLDE) and have the close trend with FPPE efficiency within the statistical uncertainty. The results show the FPPE and PCLDE change because the variation in the degree of defects concentration and both the FPPE and PCLDE depend on the number and energy of subcascades beyond the threshold energy of subcascades.

## 2. Theoretical Methods

### 2.1 Molecular dynamics

The program used here was a development of the MOLDY code of Finnis [20]. The simulation cells were first relaxed for 50 ps at 0kbar, 100 K, and 300 K. Then the recoil near the center was given an energy ranging from 0.5 to 250 keV in a random direction. The crystal was then allowed to evolve for 60 ps. AMS potential [17] were used and a variable time step and periodic boundaries but no electronic stopping were used. The crystal size was 9,826,000 atoms ($\equiv (170a_0)^3$). We ran six simulations to generate meaningful cascade statistics for every simulation with different temperatures and energies of PKA. Wigner-Seitz (WS) cells around the perfect lattice positions were employed to find the Frenkel pairs produced in the cascade simulations and we check the number of atoms inside this volume (no atom = vacancy, V: more than one atom = self-interstitial atom, SIA) [4].

**2.2 Pair analysis technique**

To analyze the structural changes of materials, we generalized a technique first used by Blaisten-Farojas [18] to decompose the first two peaks of the pair correlation function, which is widely used to study the microstructure in the noncrystalline state [21-23]. In this technique, pairs of atoms are characterized with four indices and are classified by (i) whether or not they are near-neighbors, (ii) the number of near-neighbors they have in common, and (iii) the near-neighbor relationships among the shared neighbors. Two atoms are said to be near-neighbors if they are within a specified cutoff distance of from each other. We used a cutoff distance of $1.207a_0$ in our analyses. In bcc perfect lattices, the indices are 1441 and 1661, shown in Figs. 1(a)-1(d), and their proportion is 3:4. For the 9,826,000 atoms, there are 3,930,402 1441 and 5,240,536 1661 pairs, respectively. The 1441 and 1661 pairs decrease with increase in damages, which can reflect information on damaged perfect lattices in the system. Thus, we get the PCLDE ($\eta_{pair}$) or the total 1441 and 1661 pairs damaged in the perfect lattices produced per cascade, $\eta_{pair} = v_{pd}/v_{NRT}$, that remain in the crystal when the cascade process is finished. Here, $v_{pd}$ is the total damage number of 1441 and 1661 pairs at the end of the simulations, which equal the total number of 1441 and 1661 pairs in perfect lattices minus the rest in the system.

## 3. Results and Discussion

In Fig.2 we analyze the FPPE and PCLDE versus the MD cascade energy when the cascade process is finished. The convention of assigning triangles and squares to results for simulations at 100 and 300 K, respectively, reveals that the simulation temperature has little effect, at least with this potential, in which FPPE and PCLDE seem to decrease slightly with increasing temperature at lower energies. The convention of assigning solid and hollow to results for simulations at FPPE and PCLDE, respectively, reveals that they have similar trends with some differences in temperature at high energies. The line drawn through the data in each figure, which is FPPE at 100 K, FPPE at 300 K, PCLDE at 100 K and PCLDE at 300 K in turn, is a nonlinear least square fit to the data using the following function:

$$\eta_{FP} = 0.58041 E_{MD}^{(-0.24423)} + 1.25*10^{-3} E_{MD},  \quad (3)$$

$$\eta_{FP} = 0.43288 E_{MD}^{(-0.24371)} + 1.34*10^{-3} E_{MD},  \quad (4)$$

$$\eta_{pair} = 97.70259 E_{MD}^{(-0.28736)} + 2.3476*10^{-1} E_{MD},  \quad (5)$$

$$\eta_{pair} = 93.98643 E_{MD}^{(-0.28690)} + 1.7248*10^{-1} E_{MD},  \quad (6)$$

where $E_{MD}$ is in keV.

From the above formulas and Fig. 2 we can find that the trend of these curves is similar to Eqs. (2) and can be divided into two stages: branches of cascades overlapping and subsidiary cascades production. In the first phase, described by the first term of the equations, the cascade energy is below $E_d$, and these curves decrease with MD cascade energy in the power law. This is due to an increasingly serious overlap in branches of cascade as the MD cascade energy increases, which reflects the concentration trend of defects. The cascade energy is up to about $E_d$ where FPPE and PCLDE are at the minimum value. In the second phase, the cascade energy is from 20 to 250 keV. From Figs. 3(a) and 3(b) we see that in the process of MD simulation of 250 keV cascade, though atoms pass block and come back due to periodic boundaries, the regions of defects production are quite a long way off in 0.5 ps, at which damage is maximum. So the result of MD simulation of 14.1 MeV neutron is correct. FPPE and PCLDE slightly increase with MD cascade energy in the phase. This is due to the production of many subcascades after the MD cascade energy increase to the threshold energy of subcascades, as shown in Figs. 4(a)-4(d). Since the MD cascade energy is directly proportional to the number of subcascades [23] ascribed by the second terms in Eqs. (3), (4), (5), and (6), the number of subcascades plays an important role in the increase of FPPE and PCLDE as MD cascade energy increases. Only when energy of every subcascade becomes $E_d$, FPPE and PCLDE will reach the value that appears in $E_d$, although this is impossible. Thus, in $E_d$, FPPE and PCLDE reach their minimum. When the energy is beyond $E_d$, the curves will depend on two values: the number of subcascades that make the curves increase, and the energy of single subcascade that make the curves decrease.

## 4. Conclusions

We perform the cascade simulation of MD with 14.1 MeV neutron. By using the pair analysis technique, we can acquire information on damaged perfect lattices in the system. Qualitatively analyzing about the FPPE and PCLDE versus the MD cascade energy, we find that FPPE and PCLDE reflect the degree of concentration of defects in the system, which depend on the number and energy of subcascades. The PCLDE is clearly defined and can reflect the information of defects from another perspective.

**Acknowledgements**

This work was supported by the Innovation Program of Chinese Academy of Sciences (Grant No.: KJCX2-YW-N35) and the National Magnetic Confinement Fusion Program (Grant No.: 2009GB106005), and by the Center for Computation Science, Hefei Institutes of Physical Sciences.

**References**

[1] A F Calder, D J Bacon. 1993, J. Nucl. Mater., 207: 25

[2] M W Finnis, J E Sinclair. 1984, Philos. Mag. A., 50:45

[3] M W Finnis, J E Sinclair. 1986, Philos. Mag. A., 53:161

[4] L Malerba. 2006, J. Nucl. Mater., 351:28-38

[5] C Björkas, K Nordlund. 2007, Nucl. Instr. and Meth. In Phys. Res. B., 259:835-860

[6] D J Bacon, A F Calder, F Gao, et al. 1995, Nucl. Instr. and Meth. In Phys. Res. B., 102:37-46

[7] M T Robinson, I M Torrens.1972, Phys. Rev. B., **9**: 5008

[8] I.M. Torrens, M.T. Robinson. 1972, Radiation-Induced Voids in Metals. US Atomic Energy Commission, Washington, DC: J.W. Corbett, L.C. Ianniello

[9] R E Stoller, L R Greenwood. 1999, J. Nucl. Mater., 271-272:57-62

[10] W J Phytian, A J E Foreman, R E Stoller, et al. 1995, J. Nucl. Mater., 223:245

[11] C H M Broeders, A Yu, Konobeyev. 2004, J. Nucl. Mater., 328:197

[12] M A Kirk, L R Greenwood. 1979, J. Nucl. Mater., 80:159

[13] P Jung. 1981, Phys. Rev. B., 23:664

[14] C Jaouen, J P Riviere, C Templier, et al. 1985, J. Nucl. Mater., 131:11

[15]  S Takamura, T Aruga, K Nakata. 1985, J. Nucl. Mater., 136:159

[16]  G Wallner, M S Anand, L R Greenwood, et al. 1988, J. Nucl. Mater., 152:146

[17]  G J Ackland, M I Mendelev, D J Srolovitz, et al. 2004, J. Phys.: Condens. Matter., 16:S2629

[18]  Blaisten-Barojas. 1984, E. Kinam., 6A:71

[19]  J Dana, Honeycutt, Hans C. Andersen. 1987, J. Phys. Chem., 91:4950-4963

[20]  M W Finnis. 1988, UKAEA Harvwll Report, AERE R-13182

[21]  K Y Chen, H B Liu, X P Li, et al.1995, J. Phys. Condens. Matter., 7:2379

[22]  Xia Jun Chao, Zhu Zhen Gang and Liu Chang Song.1999. Chin. Phys. Lett., 16:850

[23] Heinisch H L, Singh B N. 1992. Phil. Mag. A.,67:407-424

E-mail address of LIU Changsong: csliu@issp.ac.cn

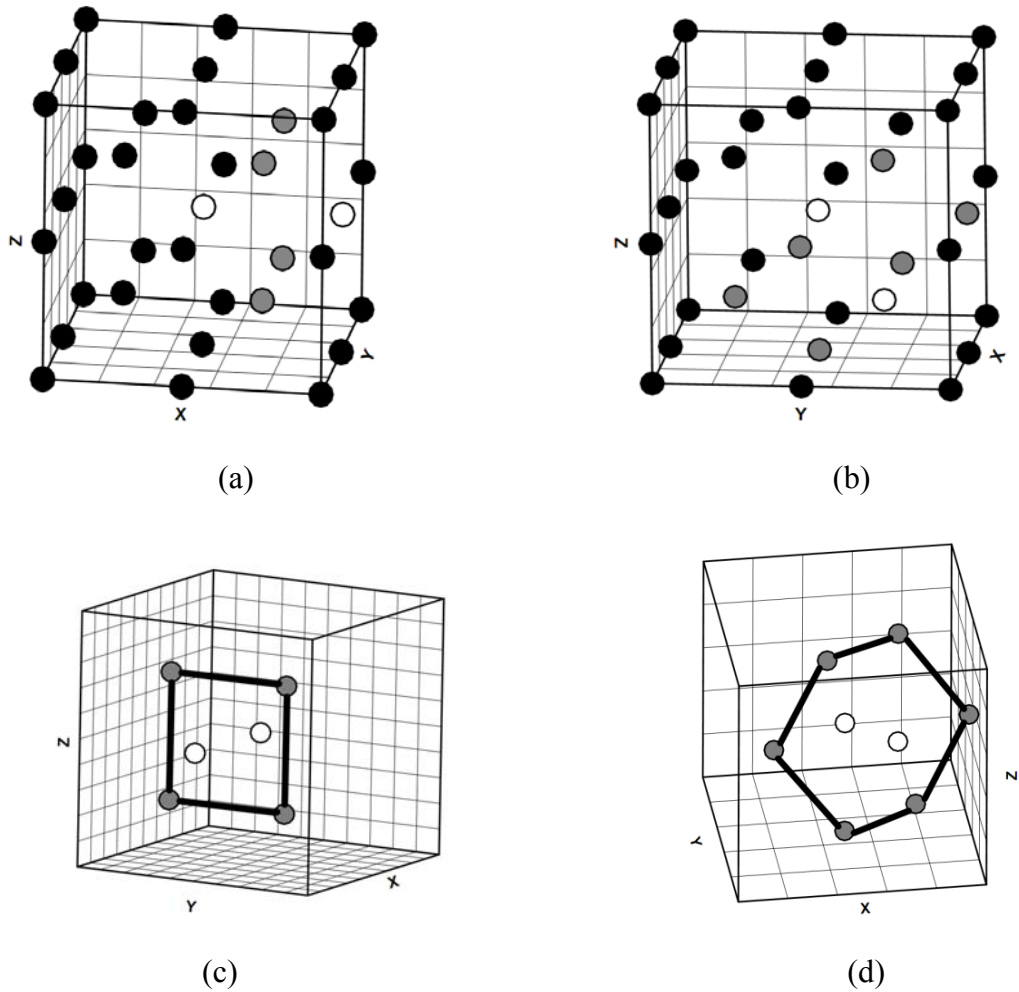

**Fig. 1** Schematic diagram of typical pairs in bcc structure, where the root pairs are denoted by open circles, and the common nearest-neighbors of the root pairs are denoted by grey circles, (a) 1441 pair in $2a_0 \times 2a_0 \times 2a_0$ block, (b) 1661 pair in $2a_0 \times 2a_0 \times 2a_0$ block, (c) 1441 pair alone, (d) 1661 pair alone.

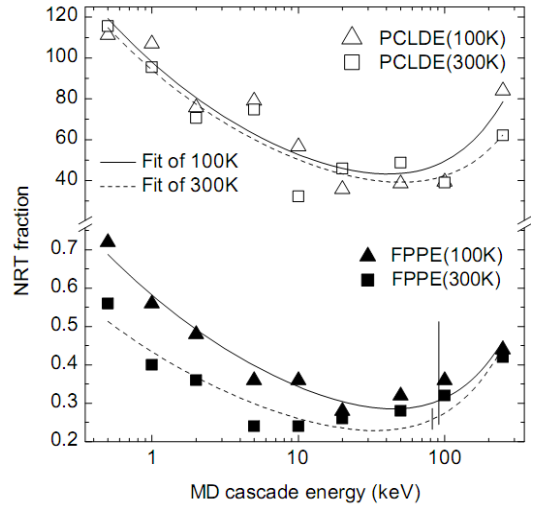

**Fig.**2 Frenkel pairs production efficiency (FPPE) and perfect crystal lattices damage efficiency (PCLDE) versus cascade energy. Full symbols are FPPE; empty symbols are PCLDE. Triangles and squares correspond to simulations at 100 and 300 K, respectively. Vertical lines are FPPE values obtained from experimentation for comparison on the right side of the figure. The solid curve describes the fit values at 100 K and the dot corresponds to the fit values at 300 K.

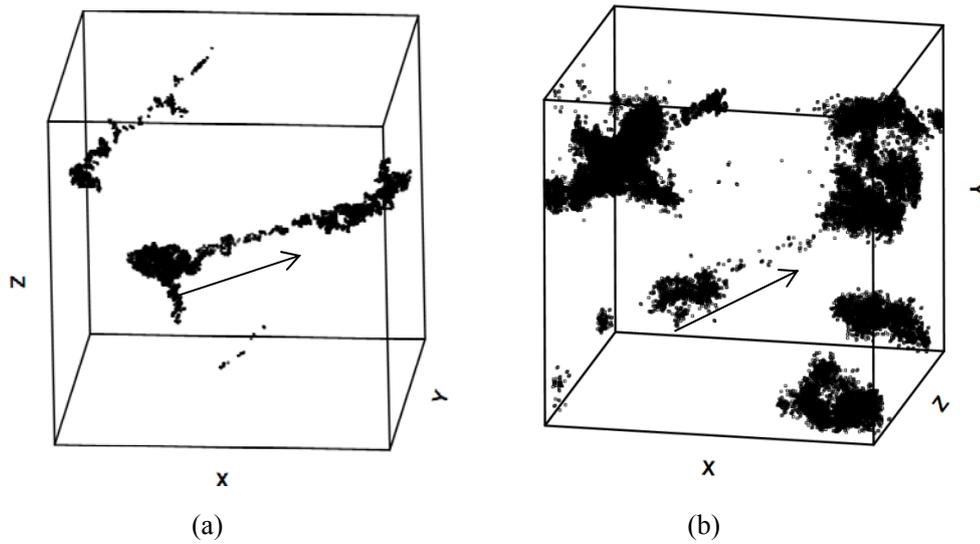

**Fig.**3 Representative snapshots of MD simulaiton of 250 keV cascades after 0.25 ps (a) and 0.5 ps (b) at 100 K. The block size is $170a_0 \times 170a_0 \times 170a_0$ and the PKA direction was ascribed by arrow. Atoms with potential energies less than some value are treated as nondefective and are not shown. As MD simulation of 250 keV cascade, although atoms pass blocks and return due to periodic boundaries, the areas of defects production are still far in 0.5 ps, at which damage is maximum. Thus, the result of MD simulation is correct.

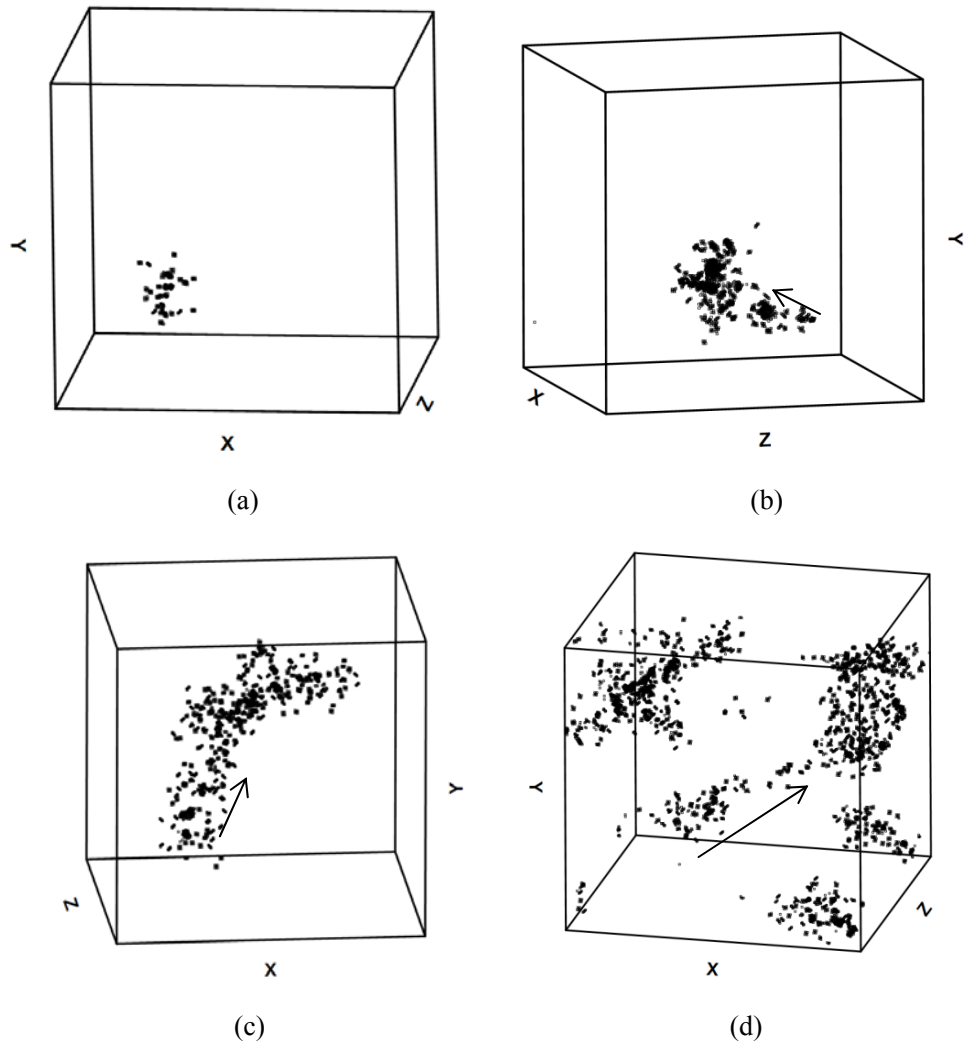

**Fig.**4 Representative snapshots of MD simulaiton of 20 keV (a) , 50 keV(b), 100 keV (c), and 250 keV (d) cascades after 60 ps at 100 K. The block size is $170a_0 \times 170a_0 \times 170a_0$, and the PKA direction is ascribed by the arrow. Atoms with potential energies less than some value are treated as nondefective and are not shown. The number of subcascades is proportionate to the MD cascade energy, which remains identical with defects scattering.